\documentclass[twocolumns]{aa}

\usepackage{graphicx}
\usepackage{natbib}
\usepackage{txfonts}
\begin{document}
   \title{A young binary Brown Dwarf in the R-CrA star formation region} 

\authorrunning{Bouy et al.}
\titlerunning{A young binary Brown Dwarf in the R-CrA region}

   \author{H. Bouy \inst{1,2,5}, W. Brandner \inst{3}, E.L. Mart\'\i n \inst{4}, X. Delfosse \inst{5}, F. Allard \inst{6}, I. Baraffe \inst{6}, T. Forveille \inst{5,7}, R. Demarco \inst{2,8} }

   \offprints{H. Bouy}

   \institute{Max-Planck  Institut f\"ur extraterrestrische Physik, Giessenachstra\ss e 1,  D-85748 Garching bei M\"unchen, Germany\\
                \email{hbouy@mpe.mpg.de}
        \and European Southern Observatory, Karl Schwarzschildstra\ss e 2, D-85748 Garching bei M\"unchen, Germany\\
	\and Max-Planck Institut f\"ur Astronomie, K\"onigstuhl 17, D-69117 Heidelberg, Germany\\
		\email{brandner@mpia.de}
	\and Instituto de Astrofisica de Canarias, 38200 La Laguna, Spain \\
		\email{ege@ll.iac.es}
	\and Laboratoire d'Astrophysique de l'Observatoire de Grenoble, 414 rue de la piscine, F-38400 Saint Martin d'H\`ere, France\\
		\email{Xavier.Delfosse@obs.ujf-grenoble.fr}
	\and Centre de Recherche Astronomique de Lyon (UML 5574), Ecole Normale Sup\'erieure, 69364 Lyon Cedex 07, France\\
		\email{fallard@ens-lyon.fr}
		\email{ibaraffe@ens-lyon.fr}
	\and Canada-France-Hawaii Telescope Corporation, PO Box 1597, Kamuela Hi-96743, USA\\
		\email{forveille@cfht.hawaii.edu}
	\and Institut d'Astrophysique de Paris, 98bis, Bd Arago,  F-75014 Paris, France \\
		\email{demarco@iap.fr}
      }

   \date{Received 30/12/2003 ; accepted 29/04/2004...}

   \abstract{
   We present imaging and spectroscopic observations with HST~(WFPC2, ACS/HRC and STIS), VLT~(FORS2) and Keck~(HIRES) of the dM8 ultra-cool dwarf \object{DENIS-P~J185950.9-370632}, located in the R-CrA region. The presence of lithium absorption at 670.8~nm and the strong H$\alpha$ emission indicate a young age and a sub-stellar mass. Our diffraction-limited images resolve a companion at the separation limit of HST/ACS~($\sim$ 0\farcs06). The 2.1 mJy flux in the LW2 filter (5.0-8.5$\mu$m) of the Infrared Space Observatory \citep{1999A&A...350..883} likely corresponds to an infrared excess, suggesting the presence of circumstellar material. Proper motion and photometric measurements, as well as the H$\alpha$ activity, confirm membership in the R-CrA star forming region. If confirmed by further observations, DENIS-P~J185950.9-370632 would be the first accreting sub-stellar multiple system observed to date.

   }

\maketitle

\emph{Note: Version with higher resolution figures available on ftp://ftp.mpe.mpg.de/people/hbouy/publications}
%

\section{Introduction: Disks and Binaries at the sub-stellar limit}
Over the last few years large numbers of sub-stellar objects have been discovered as a result of the new sky surveys \citep[2MASS, DENIS, SDSS, see respectively ][]{2MASS,1997Msngr..87...27E,2000AJ...129..1579},  making detailed studies of their physical properties possible. One of the most debated question is how do brown dwarfs originate? Are they ejected stellar embryos \citep{2001AJ....122..432R,2002MNRAS.332L..65B,2003MNRAS.342..926D}? Do they instead form isolated, like ordinary stars, due to the fragmentation of collapsing molecular cloud clumps \citep{Protostars..Planets..IV, 1987ARA&A..25...23S}, or do they form more like ``planets'', in gravitationally unstable regions of circumstellar disks \citep{2001MNRAS.325..221P}? The existence of brown dwarf binaries or the presence of disks are essential diagnostics of these formation scenarii. \\

Several dozens very low mass multiple systems have been found both in the field and in open clusters. In the field the binary fraction for separations between 5 and 20~A.U is $\sim$15\% \citep{2003AJ....126.1526B, 2003ApJ...586..512B, 2003ApJ...587..407C, 2003AJ....125.3302G}. A recent study of colour-magnitude diagrams in the Pleiades suggests that the overall binary fraction in that open cluster might be as high as $\sim50^{+11}_{-10}\%$ \citep{2003MNRAS.342.1241P}, although recent imaging surveys  \citep{2003ApJ...594..525M, 2000MNRAS.316..827R,1999AJ....118.2460B} found at most $\sim$12\% visual binaries (separations greater than $\sim$7.5~A.U). Surveys looking for disks have also found large numbers of disks around very low mass objects in star forming regions and clusters \citep[see e.g][, and references therein]{2003ApJ...590L.111P, 2002ApJ...571L.155T, 2002ApJ...578L.141J, 2001A&A...376L..22N, 2000ApJ...545L.141M}. From near-infrared observations, \citet{2003AJ....126.1515J}  estimated initial frequencies of disk around brown dwarfs ranging from 33\% to 70\% for the $\rho$--Oph, IC~348, Taurus, Cha~\textsc{I}, Upper Sco, $\sigma$--Ori and TW Hya star forming regions, and \citet{2001ApJ...558L..51M} measured over 50\% in the Trapezium Cluster. Although this suggests that brown dwarfs originate like stars more than like ejected embryos (which should have no disk), our understanding of the formation processes is far from complete. The truth might well be a combination of the different processes that have been proposed.

In the course of a large search for field late-M and L dwarfs, (Delfosse et al. in prep.) we have selected for follow-up spectroscopy all objects in 5700 squares degrees of the DENIS survey \citep{1997Msngr..87...27E} at high galactic latitude that have $I-J>3.0$~mag. A few fields at intermediate galactic latitudes (15\degr to 20\degr) overlap some nearby star forming regions. In the Corona Australis (CrA) molecular cloud complex, we selected one red object, with colours typical of very late M-dwarfs, DENIS-P~J185950.9-370632. The CrA region \citep[see e.g][]{2000A&AS..146..323N} is one of the nearest regions \citep[$\sim$130~pc; ][]{1981AJ.....86...62M, 1998AJ....115.1617C} where intermediate and low-mass stars form. Its estimated age ranges between 0.5--10~Myr \citep{1973ApJ...179..847K,1999AJ....117.2381P, 2000A&AS..146..323N}. The complex contains a dark cloud close to the R--CrA star, reflection nebulae, and several embedded infrared (IR) sources, some of which are IR class I objects \citep{1987ApJ...312..788A,1994ApJ...420..837A}. A number of T Tauri stars are associated with the CrA cloud \citep[see][]{2000A&AS..146..323N} and the faintest of those are brown dwarfs candidates or objects in the transition region between low mass stars and brown dwarfs \citep{1997AJ....114.2029W, 2001A&A...380..264F} with spectral types around M6.

In this paper, we present the results of optical and near-infrared imaging and spectroscopy of DENIS-P~J185950.9-370632. In section \ref{obs}, we will describe the observations, then in section \ref{analysis} we will give the results of these observations, and in section \ref{discussion} we will analyse these results  in the context of recent models of formation and evolution.


\section{Observations \label{obs}}

\subsection{DENIS observations \label{denis1859}}
The DENIS observations are carried out on the ESO 1~m telescope at La Silla. Dichroic beam splitters separate three channels, and a focal reducing optics provide scales of 3\farcs0 per pixel on the 256$\times$256 NICMOS3 arrays used for the two infrared channels, and 1\farcs0 per pixel on the 1024$\times$1024 Tektronix CCD detector of the I channel.  The image data were processed with the standard DENIS software pipeline \citep{DENIS_Reduction} at the  Paris Data Analysis Center (PDAC). Source extraction and  photometry are performed at PDAC, using  a space-varying kernel algorithm \citep{2000A&AS..144..363A}. When searching very rare objects in a large data base, the first challenge is artefact rejection. We use a set of morphological parameters, based on the correlation between PSF model and the object profile and on the consistency between several aperture magnitudes (Delfosse et al. in prep.).

\begin{figure*}
   \centering
   \includegraphics[width=\textwidth]{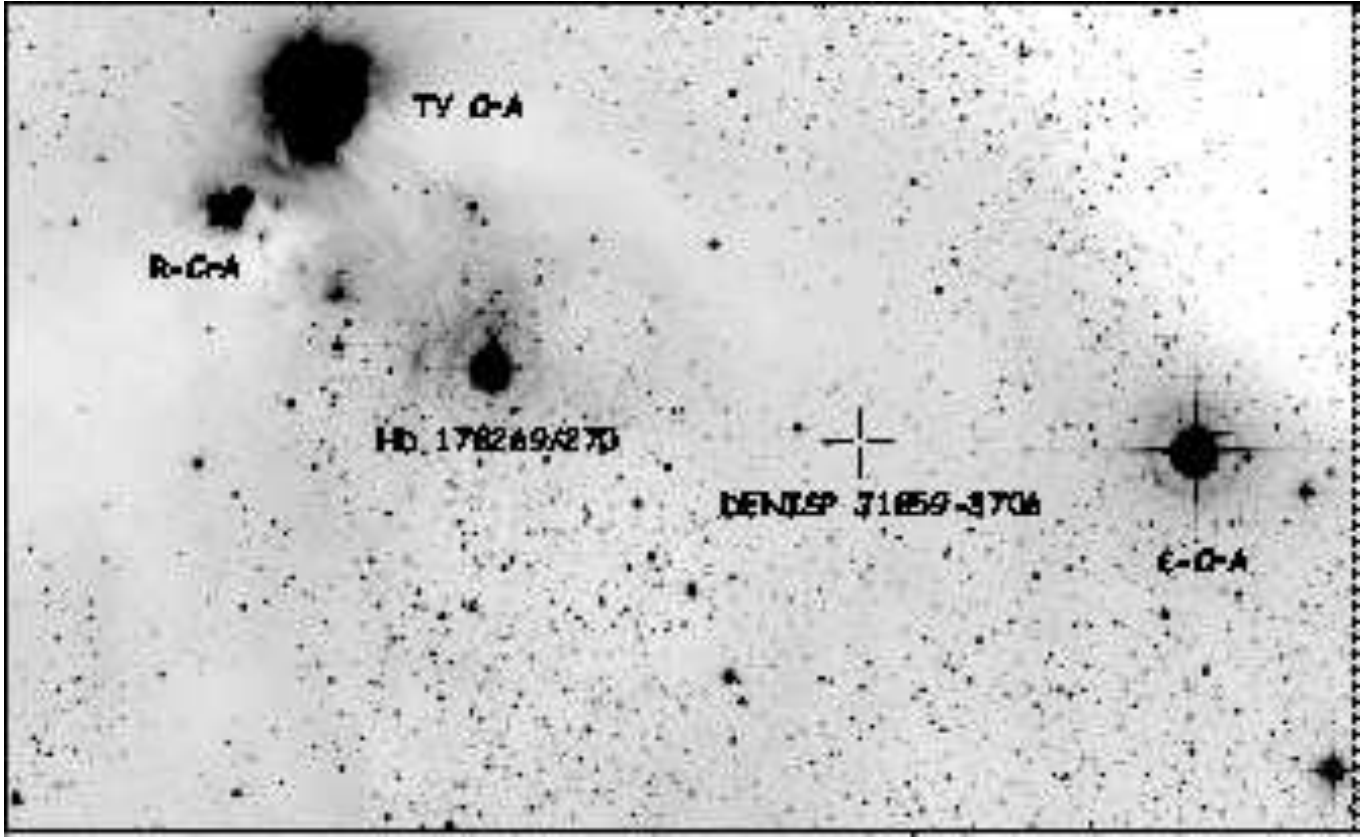}
   \caption{The environment of DENIS-P~J185950.9-370632 in the CrA molecular cloud complex. The cross marks the position of the DENIS objects on this DSS1/STSci J plate. The field of view is 50\arcmin $\times$ 35\arcmin. North is up and East is left.}
   \label{position}
   \end{figure*}

\begin{figure*}
   \centering
   \includegraphics[width=\textwidth]{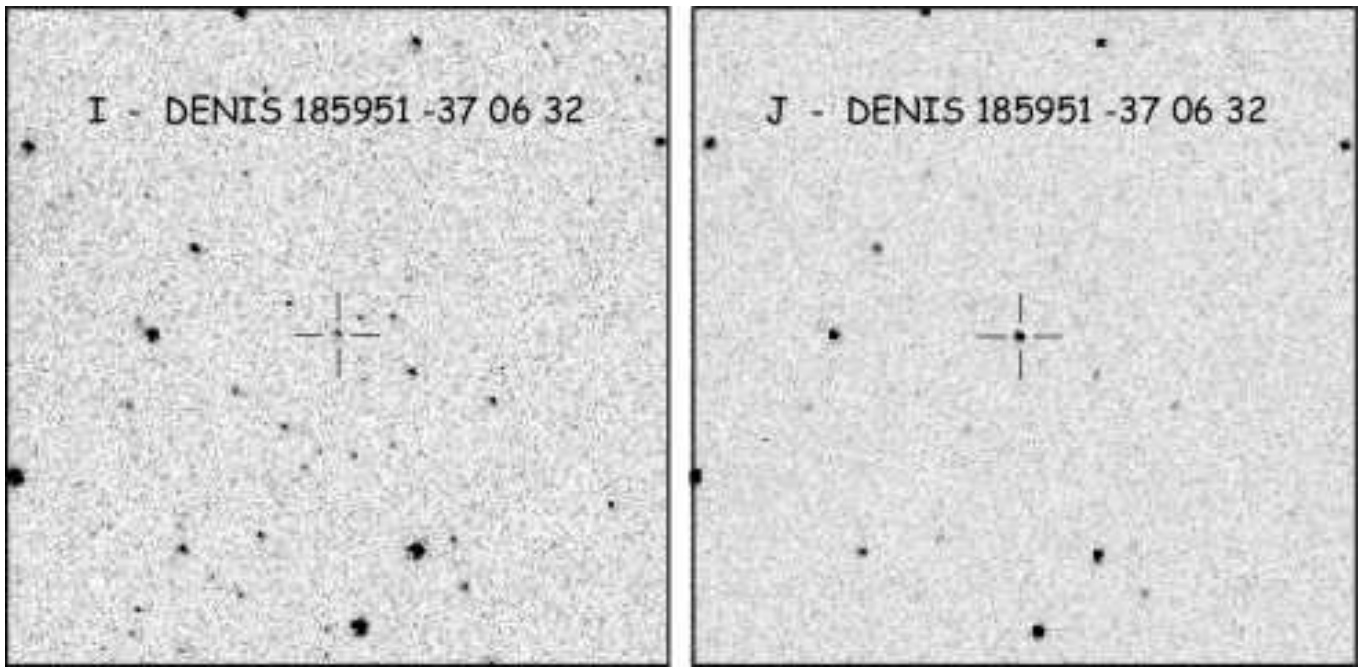}
   \caption{Finding charts of DENIS-P~J185950.9-370632 in the I and J filters (DENIS images). The field of view is 3.5\arcmin $\times$ 3.5\arcmin, North is up and East is left.}
   \label{finding_charts}
   \end{figure*}

DENIS-P~J185950.9-370632, was selected as a candidate very low mass star or and brown dwarf due to its red colors ($I-J=3.0\pm0.1$~mag). It is located in the R-CrA complex between the TY-CrA and $\epsilon$-CrA stars (see Figure \ref{position}), and very close to the core of the very dense molecular cloud \citep[$A_V~\sim~45$~mag;][]{1992ApJ...397..520W}. Locally the absorption is low, however (see Section \ref{spt}). Figure \ref{finding_charts} shows finding charts in I and J.

The apparent magnitude of DENIS-P~J185950.9-370632 is consistent with a $\sim$5~Myr late-M dwarf at the distance of the CrA complex (see section \ref{membership}), which makes it an excellent brown dwarf candidate.

\begin{figure*}
   \centering
   \includegraphics[width=\textwidth]{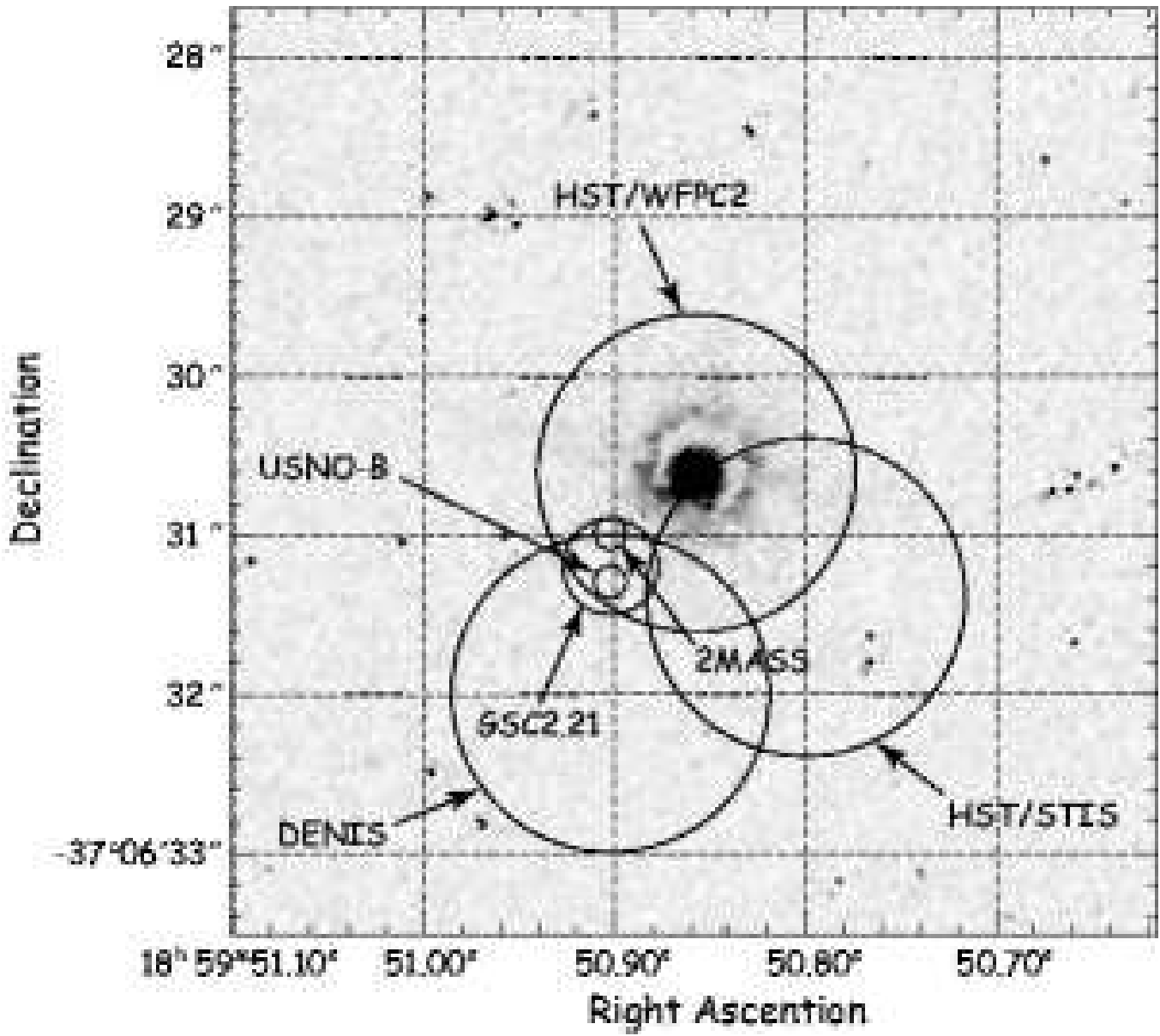}
      \caption{HST-WFPC2/PC image obtained in the F675W and F814W filters (composite image) on 2000 September 12th. The over-plotted circles represent the position of DENIS-P~J185950.9-370632 at different epochs and with different instruments. The radius of each circle corresponds to the uncertainty of that measurement. Although the different measurements are not fully consistent, no other source than DENIS-P~J185950.9-370632 can be associated in the image.}
         \label{astrometry}
   \end{figure*}

In addition to the DENIS survey, DENIS-P~J185950.9-370632 has been detected in several sky surveys, such as the USNO-B1.0 catalogue \citep[][, where it is reported as \object{USNO-B1.0 0528-0926219}]{2003AJ....125..984M}, the GSC2.21 catalog (reported as \object{GSC2 S33202002902}), and the 2MASS survey, where it is reported as \object{2MASSW18595094-3706313}. Table \ref{astrophotometry} and Figure \ref{astrometry} summarize these astrometric and photometric measurements.

\subsection{Observation Summary}

Table \ref{observations} gives an overview of all the observations we have been conducting or retrieved from the archives. We obtained high angular resolution imaging using the \emph{Hubble Space Telescope} (HST). Spectroscopy was obtained at both high (Keck-HIRES) and low (VLT-FORS2 and HST-STIS) spectral resolution. Finally, we retrieved ISO archival data in which DENIS-P~J185950.9-370632 was detected.

\subsubsection{Imaging}
We observed DENIS-P~J185950.9-370632 twice at high angular resolution using the HST, first with the Planetary Camera (PC) of WFPC2 \citep{WFPC2_DATA_HANDBOOK} and second with the High Resolution Channel (HRC) of ACS \citep{ACS_DATA_HANDBOOK}. The WFPC2 data were taken in SNAPSHOT mode (program GO8720, P.I Brandner). We took single exposure for each filter (F675W and F814W), to minimise overheads and increase exposure times. This prevent automatic removal of cosmic ray hits, but we were fortunate enough that no cosmic ray fell close enough to the target that it would affect the analysis. The ACS data (program GO9451, P.I Brandner) were obtained on 2002 September 24th in CR-SPLIT mode and with a four points dithering pattern in each of the F625W, F775W, and F850LP filters. This allows correction for cosmic ray events and bad pixels. The WFPC2 and ACS data together provide accurate photometry in five different optical filters and positions at two epochs. The target-acquisition frame for the STIS spectroscopy gives a third-epoch position.

DENIS-P~J185950.9-370632 is obviously elongated on the WFPC2, ACS and STIS images (Figure \ref{contour}). PSF subtraction brings up the companion more clearly on the WFPC2 and ACS images (Figure \ref{acs_wfpc2}).

We retrieved ISO archival data of the R-CrA region obtained on 1996 April 20th \citep[TDT 15500328, CAM01,][]{1999A&A...350..883} with ISOCAM1 which include DENIS-P~J185950.9-370632. \citet{1999A&A...350..883} report a 2.1 mJy $\pm$ 0.4 mJy detection in the LW2 filter (5.0-8.5$\mu$m), at a position that is 4\farcs0 away from our coordinates for DENIS-P~J185950.9-370632. This is well within the pointing uncertainties of ISO \citep[$\sim$6\farcs0,][]{ISOCAM..DATA..HANDBOOK}, and we identify the infrared source, \object{ISO-CrA 63}, with DENIS-P~J185950.9-370632. The source is not detected in the LW3 filter (12-18$\mu$m), and we measured a 3~$\sigma$ upper limit of 1.1 mJy. No other optical/near-IR source is present in the $\sim$6\arcsec\, ISO error box.

\subsubsection{High Resolution Spectroscopy}
DENIS-P~J185950.9-370632 was observed on 30 May 2000 with the High Resolution Echelle Spectrometer \citep[HIRES;][]{1994SPIE.2198..362V} on the Keck I telescope. A slit width of 1\farcs15 and two-pixel binning in the spectral direction gave a resolving power of R=33\,000. The exposure time was 1\,200~s and the airmass 1.83. Fifteen echelle orders were recorded on the detector, covering the wavelength range from 667.1~nm to 895.0~nm, with gaps between the orders. The data were reduced following a standard procedure in the IRAF\footnote{IRAF is distributed by National Optical Astronomy Observatories, which is operated by the Association of Universities for Research in Astronomy, Inc., under contract with the National Science Foundation.} environment, including bias subtraction, flat-field correction, aperture extraction and wavelength calibration using a Th-Ar lamp spectrum.

\subsubsection{Low Resolution Spectroscopy}

We obtained long slit low resolution spectra with FORS2 at VLT on Paranal on 2002 March 25th. The VLT uses an active optics platform to achieve high quality image. The seeing conditions were excellent ($\sim$ 0\farcs64) and we obtained high quality spectra. We used a 0\farcs7 slit and the GR600I+25 and GR1200R+93 grisms, therefore covering a large wavelength range in the red part of the optical spectrum (Table \ref{observations}). The spectra were processed using a custom pipeline based on standard procedures in IRAF. First, the two-dimensional images for the two separate CCDs which make up the FORS2 detector were independently overscan-, bias- and flat field- corrected. The lower chip image was then normalised to the median value of the upper one, and the images for the two chips could then be merged using the \emph{fsmosaic} tool of the FIMS software (FORS1+2 FIMS manual, issue 2.6). The 1-D spectra were then extracted, and wavelength and flux calibrated in a standard manner, using the best calibration files provided by the VLT team.

We obtained a HST-STIS low spectral resolution observation, aiming for spatially resolved spectra of the two components. Observations occurred on 2003 June 26th. We used the G750L grating (0.525$\mu$m-1.300$\mu$m, 4.92\AA/pixel) with the 0\farcs2 slit oriented along the axis of the binary. Unfortunately, the small separation ($\sim$0\farcs060), the relatively small flux ratio ($\sim$0.3 in R and I, Section \ref{binarity}), and the relatively low S/N prevent us from properly resolving the two components (the pixel scale of STIS spectra is $\sim$0\farcs050). We processed the integrated spectrum using the recommended STSDAS tools in IRAF and the best calibration files provided by the STScI archive.

\section{Analysis \label{analysis}}

\subsection{Spectral Type and extinction \label{spt}}
\begin{figure*}
     \centering
     \includegraphics[width=\textwidth]{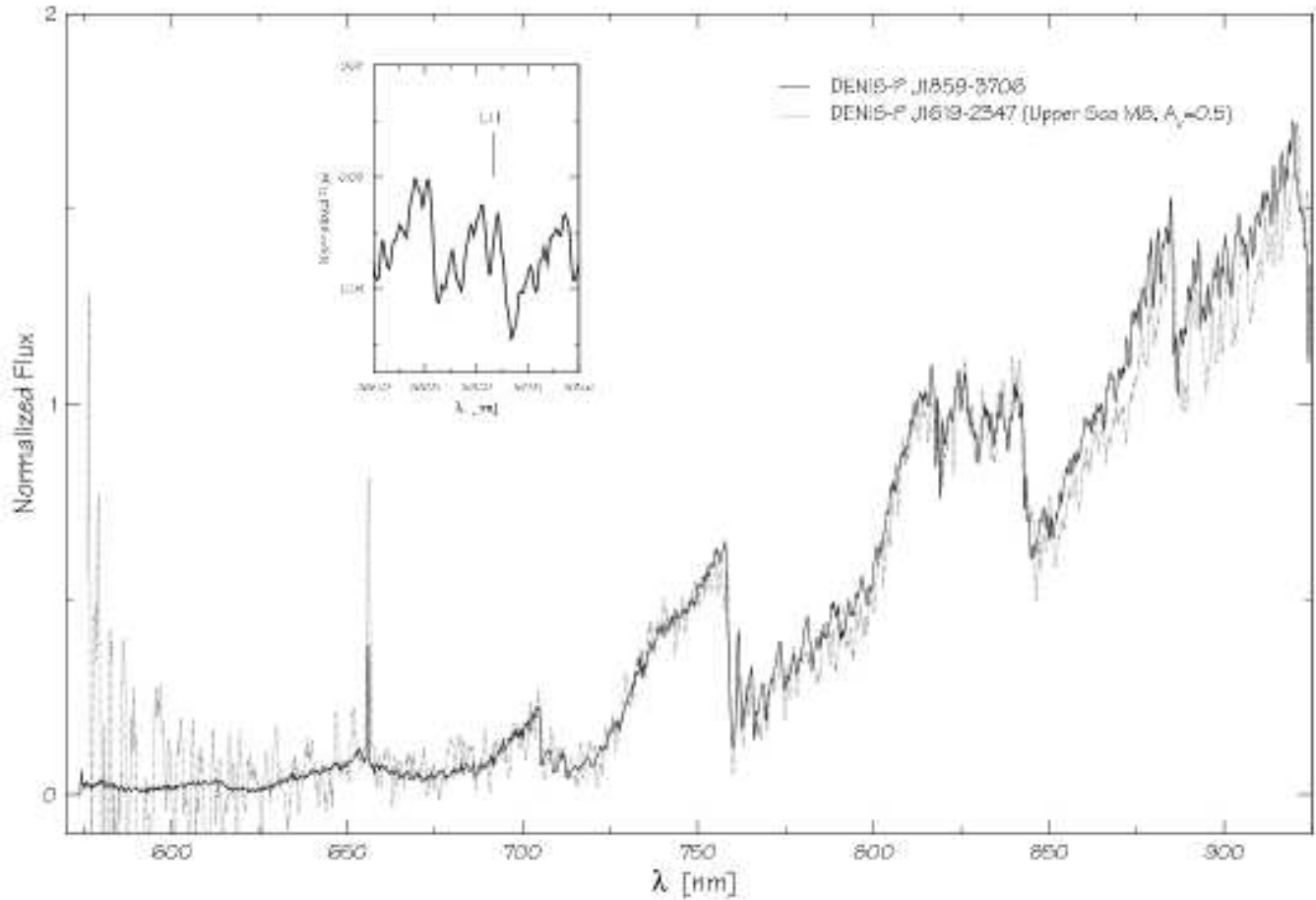} 
     \caption{Comparison of DENIS-P~J185950.9-370632 FORS2 spectrum (smoothed via a boxcar with a width of 5 pixels) with the spectrum of the Upper Scorpius member M8 dwarf DENIS-P~J161916.5-234722.9 artificially reddened with  A$_{V}$=0.5~mag. The match is very good. The inset box shows a zoom of DENIS-P~J185950.9-370632 spectrum around the Li~\textsc{I} absorption. Spectrum of  DENIS-P~J161916.5-234722.9 from \citet{2004AJ....127..449M}. Fluxes have been normalized at the pseudocontinuum level at 840.0~nm.}
     \label{denis1859_spt}
   \end{figure*}

\begin{figure*}
     \centering
     \includegraphics[width=\textwidth]{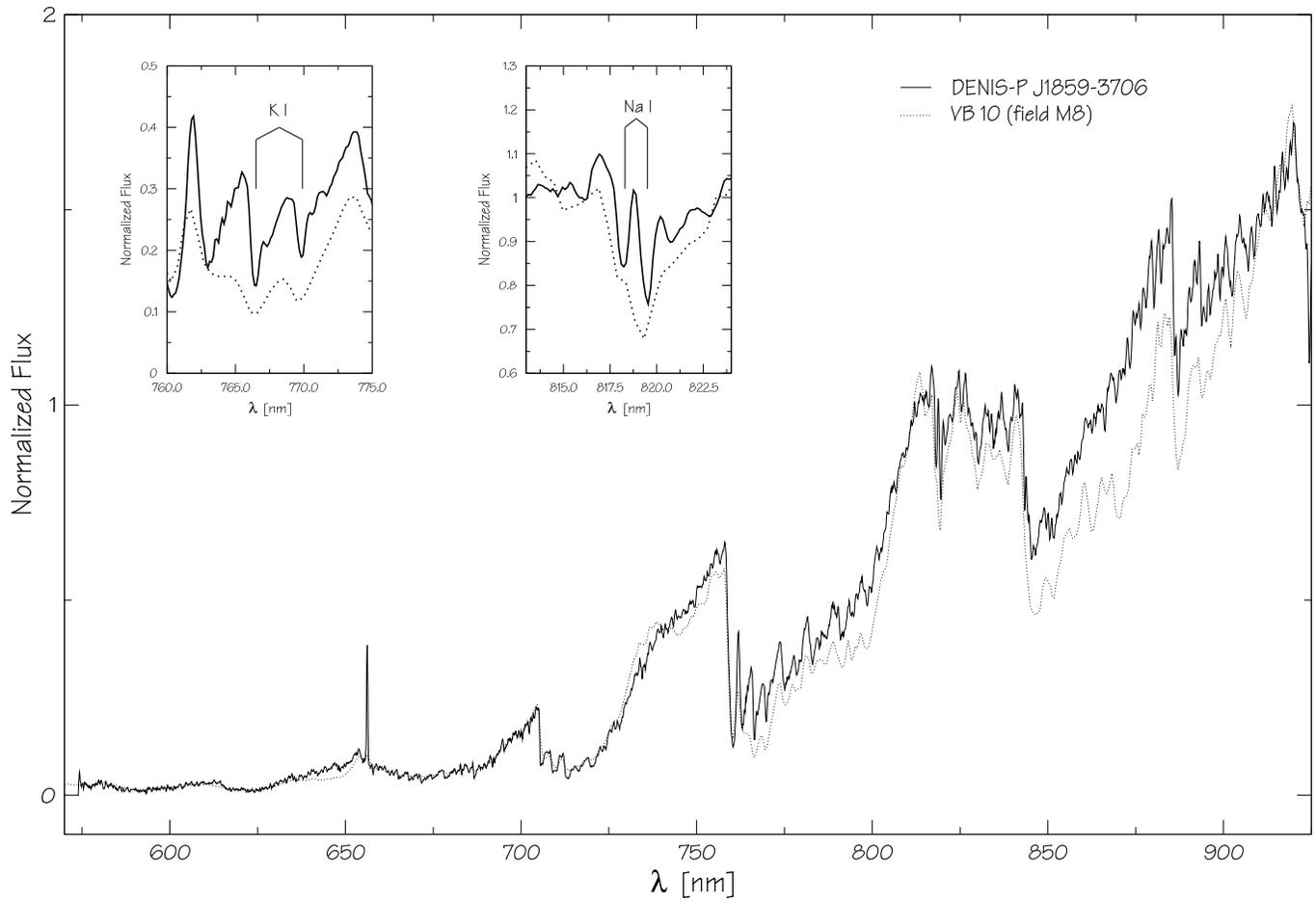} 
     \caption{Comparison of DENIS-P~J185950.9-370632 FORS2 spectrum (smoothed via a boxcar with a width of 5 pixels) with the spectrum of the field M8 dwarf \object{VB~10}, artificially reddened with  A$_{V}$=0.5~mag. The match is very good. The inset boxes show zooms of the two spectra around the K~\textsc{I} and Na~\textsc{I} doublet. It shows clearly that these two doublets are stronger in VB~10, as expected because of its higher gravity. Spectrum of VB~10 from \citet{1999AJ....118.2466M}. Fluxes have been normalized at the pseudocontinuum level at 840.0~nm.}
     \label{denis1859_vb10}
   \end{figure*}

In order to estimate the spectral type of DENIS-P~J185950.9-370632, we compare its optical spectrum to that of brown dwarfs from the field, and from the Upper Scorpius OB association \citep[hereafter USco, ][]{2003sf2a.confE.242D, 2004AJ....127..449M}. The age and distance of USco, respectively estimated at 5~Myr \citep{1999AJ....117.2381P} and 145~pc  \citep{1999AJ....117..354D}, are very close to those of R--CrA, and A$_{V}$ is close to zero, with local maxima reaching $\sim$1.0~mag \citep[see][]{1999A&A...345..965C}. Each reference spectrum was artificially reddened with different values of extinction (using the \emph{redden} task of IRAF) and then compared to DENIS-P~J185950.9-370632 spectrum. Figures \ref{denis1859_spt} and \ref{denis1859_vb10} shows the best match obtained with the USco M8 brown dwarf \object{DENIS-P~J161916.5-234722.9}, and the M8 field ultracool dwarf \object{VB~10}, both for an extinction of A$_{V}$=0.5~mag. The correlation is very good and the two results agree perfectly. One can note that the K~\textsc{I} and Na~\textsc{I} doublets are stronger in VB~10, as expected because of its higher gravity, while they are very similar in the case of DENIS-P~J161916.5-234722.9. This is another hint that DENIS-P~J185950.9-370632 is young and is likely to belong to the R--CrA association, which age is very close to that of the Usco OB association. The next best match was with \object{DENIS-P~J1556-2106} (USco, M7) for a reddening of Av=1.0. We thus adopt a spectral type of M8$\pm$0.5 for DENIS-P~J185950.9-370632 and an extinction of A$_{V}$=0.5$\pm$0.3~mag.

\begin{figure*}
     \centering
     \includegraphics[width=\textwidth]{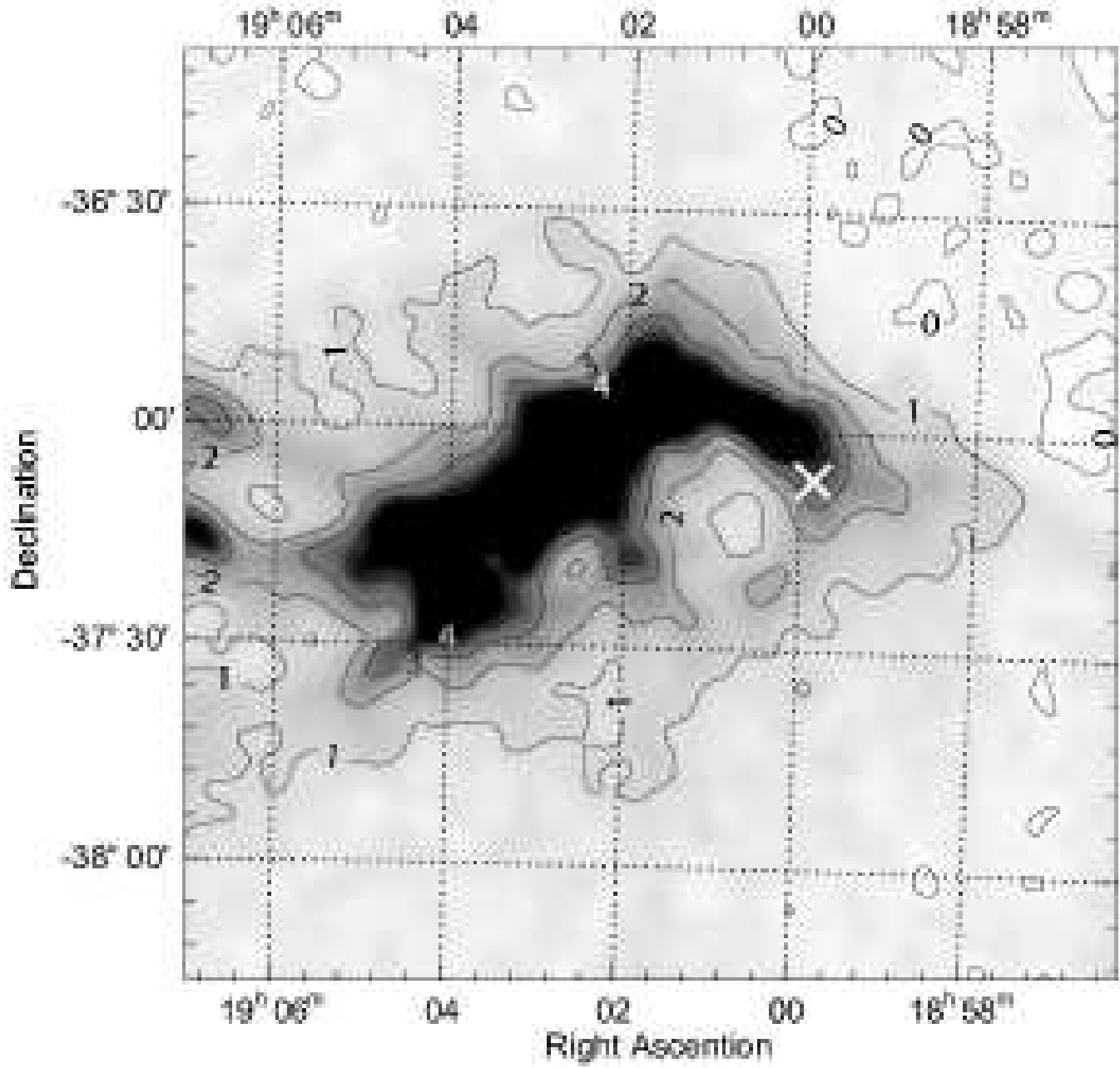} 
     \caption{Extinction map of the R--CrA Region around DENIS-P~J185950.9-370632 (indicated by white cross). Isocontours of A$_{V}$=0.0, 1.0, 2.0, 3.0 and 4.0~mag indicate the scale.}
     \label{extinction}
   \end{figure*}

We applied the \emph{NICER} (Near Infrared Colour Excess Revisited) technique \citep{2001A&A...377.1023L} on the 2MASS catalog photometry to produce an extinction map of the R--CrA region, with a 2\arcmin ~ resolution (Figure \ref{extinction}). DENIS-P~J185950.9-370632 lies next to a region of strong absorption. The mean V~band extinction in the 2\arcmin\,  pixel that contains DENIS-P~J185950.9-370632 is A$_V$=3.7$\pm$0.4~mag \footnote{The 0.4~mag uncertainty quoted here is the measurement uncertainty on the mean extinction; the standard deviation of the extinction inside the 2\arcmin\, pixel is 1.8~mag}. As explained above, the observed optical spectrum implies instead that DENIS-P~J185950.9-370632 is reddened by A$_{V}\sim$0.5~mag. This suggests that DENIS-P~J185950.9-370632 probably lies somewhat on the near side of the cloud, or in a relative gap of the patchy extinction.

  \subsection{R-CrA membership \label{membership}}
     \subsubsection{Photometric distance \label{photom_dist}}

 After correcting for A$_{V}$=0.5~mag, the magnitude and colour of DENIS-P~J185950.9-370632 are $I\sim16.8$~mag and $I-J=$2.9~mag. The DUSTY theoretical isochrones \citep{2000ApJ...542..464C} associate to those parameters a distance of $\sim$150~pc at 5~Myr and 110~pc at 10~Myr. Corrected for binarity (assuming a flux ratio $f_{2}/f_{1}=$0.3 as measured in the F814W filter, see Table \ref{result_psf_fitting} and Section \ref{binarity}), it corresponds to distances of $\sim$170~pc and 125~pc. The corresponding masses are respectively $\sim$0.025~M$_{\sun}$ and $\sim$0.035~~M$_{\sun}$. 

The late-M dwarfs sequence in Upper Scorpius \citep{2003sf2a.confE.242D, 2004AJ....127..449M} provides a sanity check: M7.5 dwarfs in USco have I between 16.0 and 17.5~mag for M6.5 to M7.5 dwarfs, indicating for DENIS-P~J185950.9-370632 a photometric distance similar to that of Upper Scorpius, whose distance is close to that of the CrA complex.

     \subsubsection{Proper motion \label{pm}}
As discussed in section \ref{denis1859}, DENIS-P~J185950.9-370632 has been detected at many epochs. Figure \ref{astrometry} shows that HST/ACS position is clearly discrepant, which we suspect is due to underestimated uncertainties on the position obtained with HST/ACS, which lies over 3-$\sigma$ away from any other. Direct comparison with the HST/WFPC2 images confirms that the pointing is correct, and we suspect that the problem arises in the astrometric processing of the ACS data by the STSDAS pipeline. The DENIS, 2MASS, HST/WFPC2, HST/STIS, GSC~2.21 and USNO-B1.0 observations all agree to within less than 1-$\sigma$ for epochs spread over 18 years, confirming that the HST/ACS measurement is most suspicious. There are unfortunately not enough other stars in the field of the HRC to perform meaningful and precise astrometric re-calibration. We derive an approximate motion for  DENIS-P~J185950.9-370632 by comparing the most precise of recent measurements, from 2MASS, with the earliest position reported in the USNO-B1.0 (with similar uncertainties, $\sim$14 years earlier). This rough estimate of the proper motion, $\mu_{\alpha}\,\cos\delta=0\farcs0 \, \mathrm{~yr}^{-1} \pm 0\farcs013\, \mathrm{~yr}^{-1}$ and $\mu_{\delta}=-0\farcs021 \, \mathrm{~yr}^{-1} \pm 0\farcs013\, \mathrm{~yr}^{-1}$, agrees within the (large) uncertainties with the \citet{2000A&AS..146..323N} value for the R-CrA region, $\mu_{\alpha}\,\cos\delta=0\farcs005 \mathrm{~yr}^{-1}$ and $\mu_{\delta}=-0\farcs027\mathrm{~yr}^{-1} $. This is consistent with membership of DENIS-P~J185950.9-370632 in the star forming region, though the significance of the result is obviously not very high.

  \subsection{Imaging: a close companion \label{binarity}}
\begin{figure*}
   \centering
   \includegraphics[width=\textwidth]{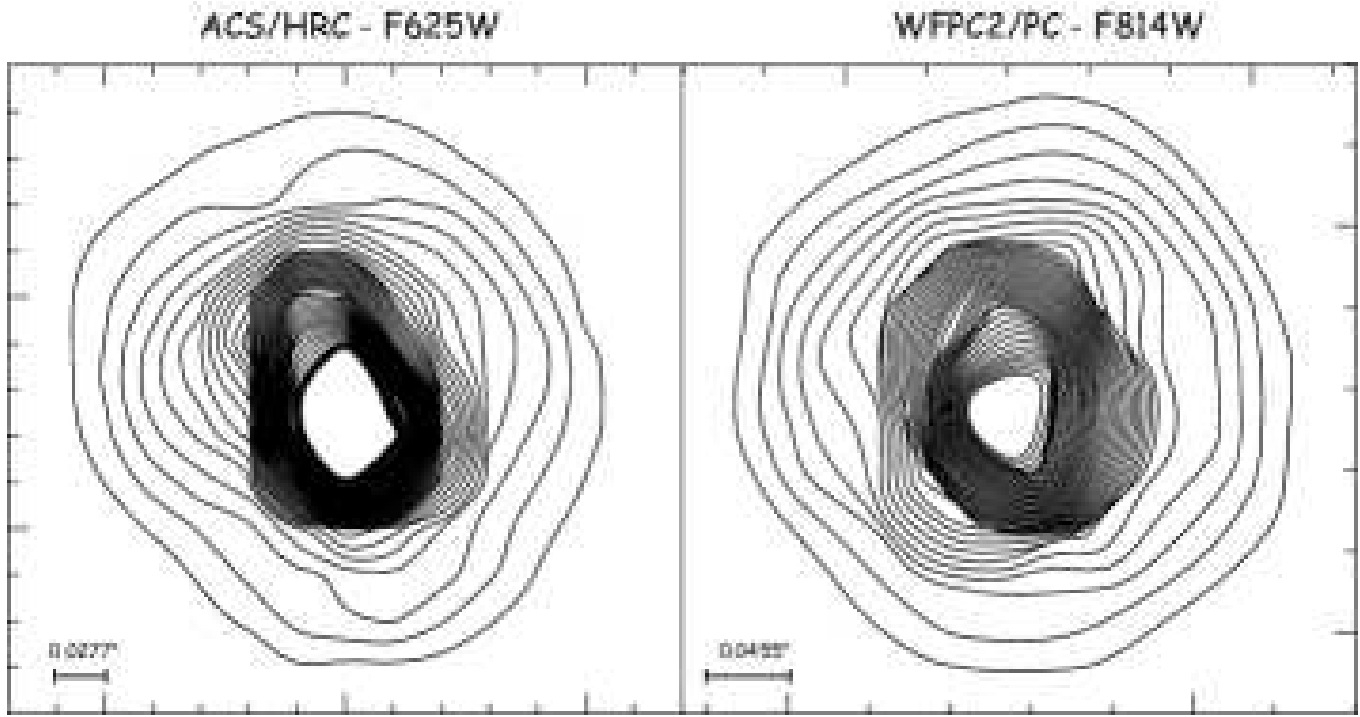}
      \caption{Contour plots of the ACS/HRC (F625W) and WFPC2/PC (F814W) images. The WFPC2/PC and ACS/HRC image have identical scales for easier comparison. The object is clearly elongated on both images, but is better resolved on the higher resolution and better sampled ACS image.}
         \label{contour}
   \end{figure*}

\begin{figure*}
     \centering
     \includegraphics[width=\textwidth]{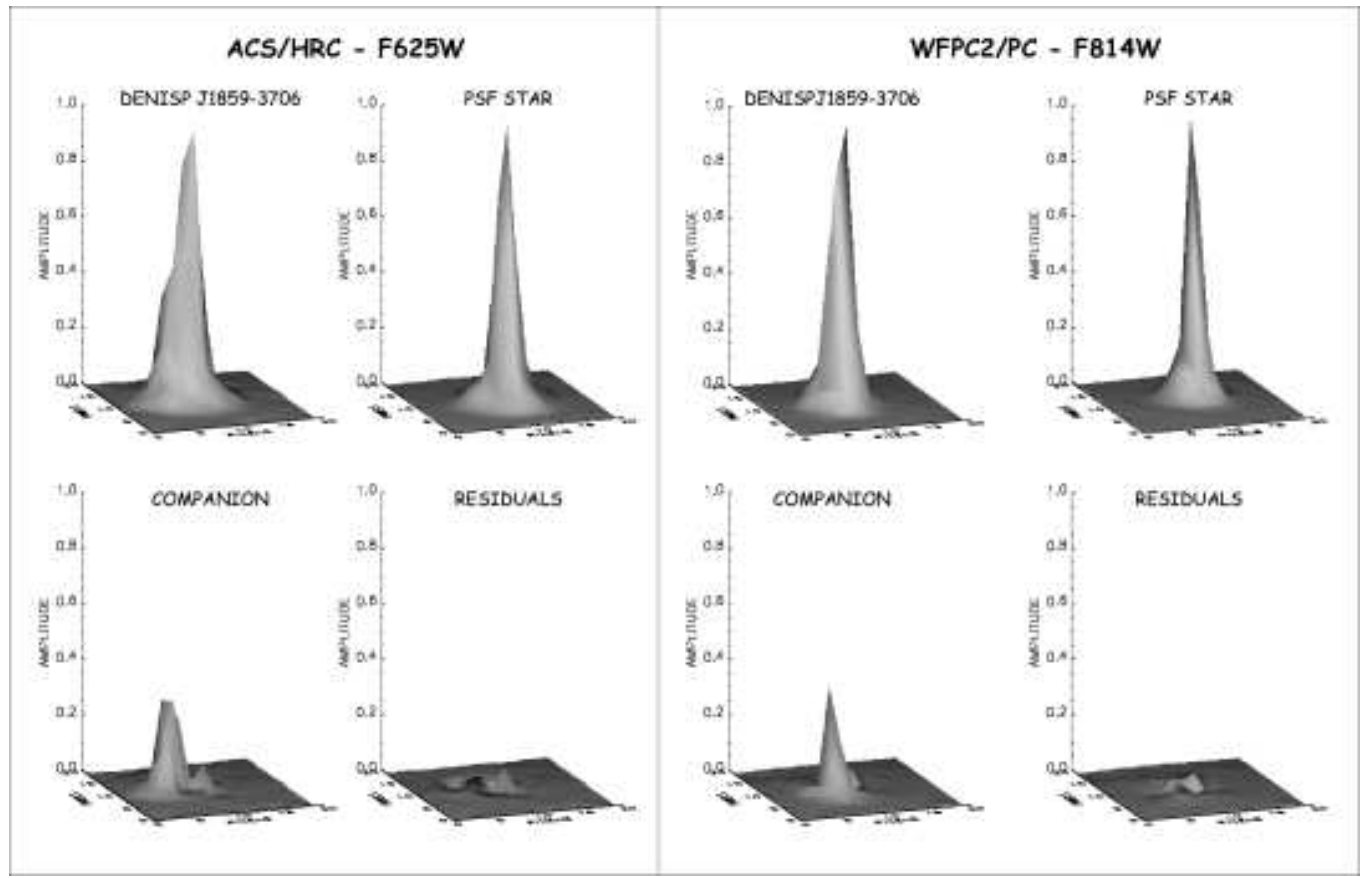} 
     \caption{Surface plots showing the results of the non-linear PSF fitting on the ACS/HRC (F625W filter) and WFPC2/PC (F814W) images. Amplitudes are normalised. The sky background has been subtracted. The figure shows the images obtained with HST/HRC and WFPC2/PC, one of the PSF stars, the companion appearing after PSF subtraction, and the residuals after subtracting the modelled binary system. The pixel scale of the ACS/HRC is almost twice finer than that of the WFPC2/PC.}
     \label{acs_wfpc2}
   \end{figure*}

As shown in Figures \ref{contour} and \ref{acs_wfpc2}, DENIS-P~J185950.9-370632 is clearly elongated in the high resolution HST images, at 2 epochs and in 6 different filters. We analysed the HST images for precise separations, position angles and flux ratios of the possible multiple system, using a custom-made program described in \citet{2003AJ....126.1526B} and adapted here for ACS/HRC. Briefly, the PSF fitting routine builds a model binary using ten different PSF stars from several ACS/HRC images, and then perform a non-linear PSF fit of the observed image to determine the best-fit values for the 3 free parameters: separation, position angle and flux ratio. \citet{2003AJ....126.1526B} discuss the uncertainties and limitations of the algorithm in detail, but slight improvements since then have led to a much better understanding of the uncertainties and systematic errors \citep[][, in prep.]{bouy2004}. 

Table \ref{result_psf_fitting} summarises the resulting best-fit binary parameters. For the three filters where the source is best resolved (F625W, F675W, F775W; see  Figures \ref{contour} and \ref{acs_wfpc2}) they are fully consistent, demonstrating that a binary star is an excellent model of the observations. The nominally discrepant parameters in the other filters are from marginally resolved images, either due to diffraction broadening at redder wavelength (WFPC2/PC F814W, ACS/HRC F850LP) or because of wider pixels (STIS/F28X50LP). They should therefore not be given much weight. Given the proper motion derived in section \ref{pm}, two independent objects should have moved apart by $\sim$0\farcs042 between the two observations. The lack of any apparent change is thus a strong indication that they form a common proper motion pair. The flux ratios indicate that the secondary is significantly fainter than the primary at all observed wavelengths.

  \subsection{Spectral Analysis \label{spectro}}
  \subsubsection{Spectral Features}
\begin{figure*}
   \centering
   \includegraphics[width=\textwidth]{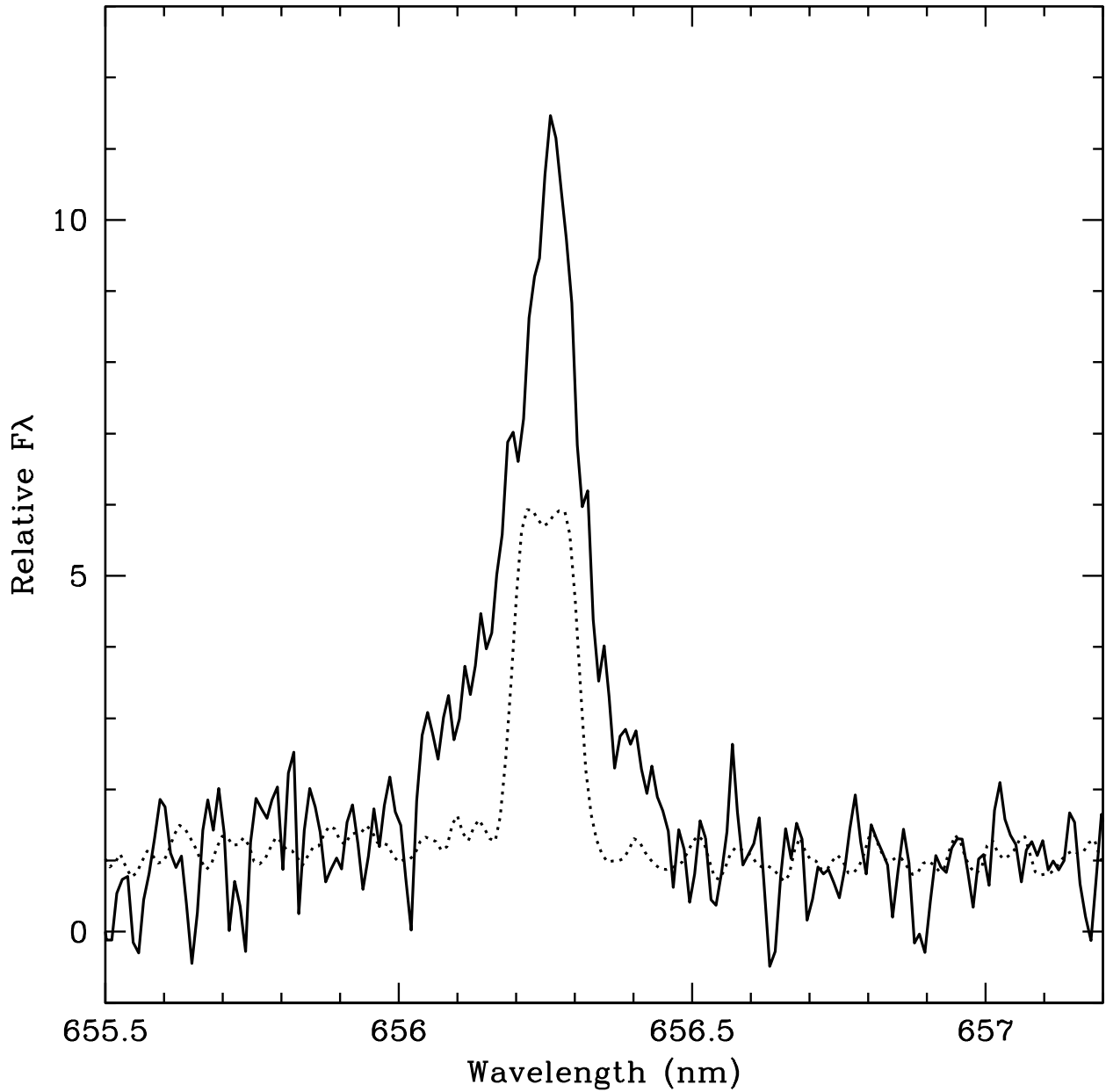}
      \caption{HIRES spectra of DENIS 1859-37 (solid line) and Gl 406 (dotted line) in the H$_\alpha$ region. Gl 406 is a typical chromospherically active M6 dwarf. DENIS 1859-37 has a much broader and asymmetric emission line, indicating that it is probably an accretor.}
         \label{halpha}
   \end{figure*}

Table \ref{spectral_features} lists the equivalent width (EW) of several spectral features measured in the different data sets. The only emission line in the HIRES spectrum is H$\alpha$. Other common emission lines such as He~\textsc{I} at 667.8~nm, O~\textsc{I} at 844.6~nm, and Ca~\textsc{ii} at 866.2~nm are not detected, with upper limits on their equivalent width (EW) below 0.5~\AA . By direct integration of the line profile (using the \emph{splot} IRAF task) we measure an H$\alpha$ equivalent width of 18$\pm$3~\AA ~ in the HIRES spectrum, 18$\pm$3~\AA ~ from the low-resolution STIS spectrum, and 17$\pm$2~\AA ~ from the FORS2 spectrum. According to \citet{Barrado...Martin} EW(H$\alpha$)=47.5~\AA ~for an M7.5 spectral type implies that the H$\alpha$ emission is caused by accretion, while weaker lines can be due to either accretion or chromospheric activity. The H$\alpha$ line strength by itself is thus here insufficient to distinguish between an accretor and a chromospherically active star. The H$\alpha$ line however is relatively broad (Figure \ref{halpha}), with a full width at 10\% of peak intensity of 205$\pm$10~km s$^{-1}$. H$\alpha$ line widths above 200~km s$^{-1}$ in brown dwarfs are due to accretion \citep{2003AJ....126.1515J}, and by this measure DENIS-P~J185950.9-370632 is slighty above the limit and is likely an accretor. If accreting, the mass accretion rate is probably rather low, given the modest H$\alpha$ strength and the lack of optical veiling.  Accretion rates below 10$^{-9}$~M$_\odot$~yr$^{-1}$ produce no measurable veiling  \citep{2000ApJ...545L.141M}. The consistency of the EW measured at the three epochs suggests that we are measuring quiescent emission rather than  variable activity. This is more consistent with steady accretion than with chromospheric activity, where strong and broad H$\alpha$ lines are only observed during outbursts. Further observations should be made in order to confirm that DENIS-P~J185950.9-370632 is accreting or not.

In the region of the Li~\textsc{I} resonance line (670.8~nm) the HIRES spectrum is rather noisy, but the line is well detected with EW(Li~\textsc{I})=0.41$\pm$0.08~{\AA}. This line also appears clearly in the FORS2 spectrum, as shown in Figure \ref{denis1859_spt}, with EW(Li~\textsc{I})=0.9$\pm$0.4~\AA. The resolution of the STIS spectrum is insufficient to isolate the Li~\textsc{I} line. The K~\textsc{I} line at 769.9~nm is also present in the HIRES and FORS2 spectra.  As already mentioned in section \ref{spt}, Figure \ref{denis1859_vb10} shows that the K~\textsc{I} and Na~\textsc{I} lines of DENIS-P~J185950.9-370632 are much narrower and weaker than in the spectrum of a field M8 dwarf (VB~10), indicating a young age.

  \subsubsection{Comparison with model spectra}
In order to estimate of the photospheric effective temperature, we compare the observed spectra with the DUSTY \citep{2001ApJ...556..357A} and NextGen \citep{1999ApJ...512..377H} atmospheric models. The synthetic spectra were smoothed with a gaussian kernel matched to the slit size and resampled to the same grid as the observed spectra. The best fit model was determined by maximizing the cross-correlation with the dereddened observed spectrum (a minimum $\chi^2$ adjustment, performed as a cross-check, gives identitical results).

The free parameters for the fit are the effective temperature and the surface gravity (the metallicity was fixed to solar metallicity; models for higher metallicities are not yet available, and lower metallicities gave significantly degraded agreement). The latest NextGen and DUSTY models give very similar surface gravities. For an age of 5~Myr and I-J$\sim$3.0, they respectively give log~$g\sim$3.8 and log~$g\sim$4.0, and for 10~Myr respectively log~$g\sim$4.1 and log~$g\sim$4.0. The available DUSTY models have a 100~K grid step, and cover surface gravities log~$g$ ranging from 3.5 to 6.0 with a 0.5 interval. We only considered models with effective temperature between 1500~K and 3000~K, and surface gravities between 3.5 and 4.5. The observed and synthetic spectra were both normalized to an integrated flux of unity prior to the analysis. 

\begin{figure*}
   \centering
   \includegraphics[width=\textwidth]{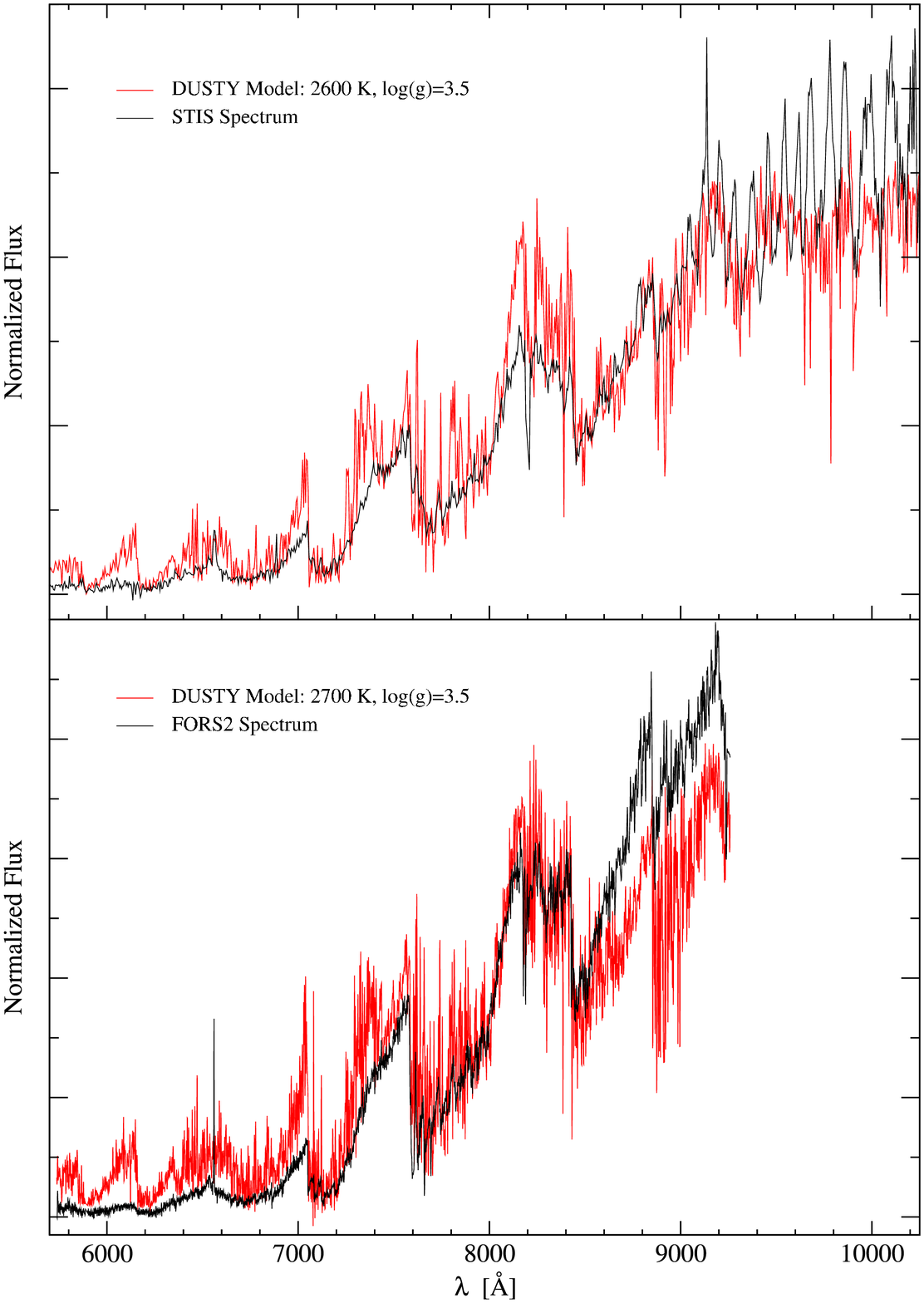}
      \caption{STIS and FORS2 spectra compared with models. Top panel: STIS spectrum of DENIS-P~J185950.9-370632 compared with a synthetic spectrum. Bottom panel: FORS2 spectrum of DENIS-P~J185950.9-370632 compared with a synthetic spectrum. The best fit is obtained for the same surface gravity (log~$g$=3.5) but slightly different effective temperatures (2600~K and 2700~K). The two results therefore agree within the 100~K temperature step of the model atmosphere grid.} 
         \label{spectra_models}
   \end{figure*}

Figure \ref{spectra_models} shows the best-fit DUSTY synthetic spectrum, over-plotted on the observed spectra. For both sets of models (DUSTY and NextGen), the best fits are obtained for T$_{eff}$=2600~K and log~$g$=3.5 (STIS spectrum), and T$_{eff}$=2700~K and log~$g$=3.5 (FORS2 spectrum). These temperatures correspond to that of the equivalent unresolved system. Attempts to account for the multiplicity by using two synthetic spectra at different effective temperatures did not produce a significantly improved fit. This is actually expected, since according to the DUSTY models the observed magnitude differences (from 0.60~mag in the HST F775W filter to 1.74~mag in the HST F625W filter) correspond to effective temperatures that differ by $\lesssim$200~K. This is close to the effective temperature resolution of our adjustment, and a dual temperature fit is therefore not warranted. The obtained effective temperature should on the other hand represent a reasonable estimate for both components.

It is important to note that the model spectra do not match the observed ones very well over the present spectral range, perhaps because of their simplistic handling of the gravitational settling of dust. They probably underestimate the strength of molecular absorption bands like TiO, VO, but of also atomic lines like K~\textsc{I} and Na~\textsc{I}~D. Since these features dominate the energy distribution in optical spectra of late M dwarfs, the parameters that we obtain most likely suffer from systematic errors. They should on the other hand be much more reliable when used in a relative sense, and compared with other analyses based on the same atmospheric models. The low surface gravity of DENIS-P~J185950.9-370632 (log~$g=3.5$) relative to field M8 dwarfs (log~$g{\sim}4.5$), in particular, is a robust result.

\section{Discussion \label{discussion}}

  \subsection{Infrared excess}
  
\begin{figure*}
   \centering
   \includegraphics[width=\textwidth]{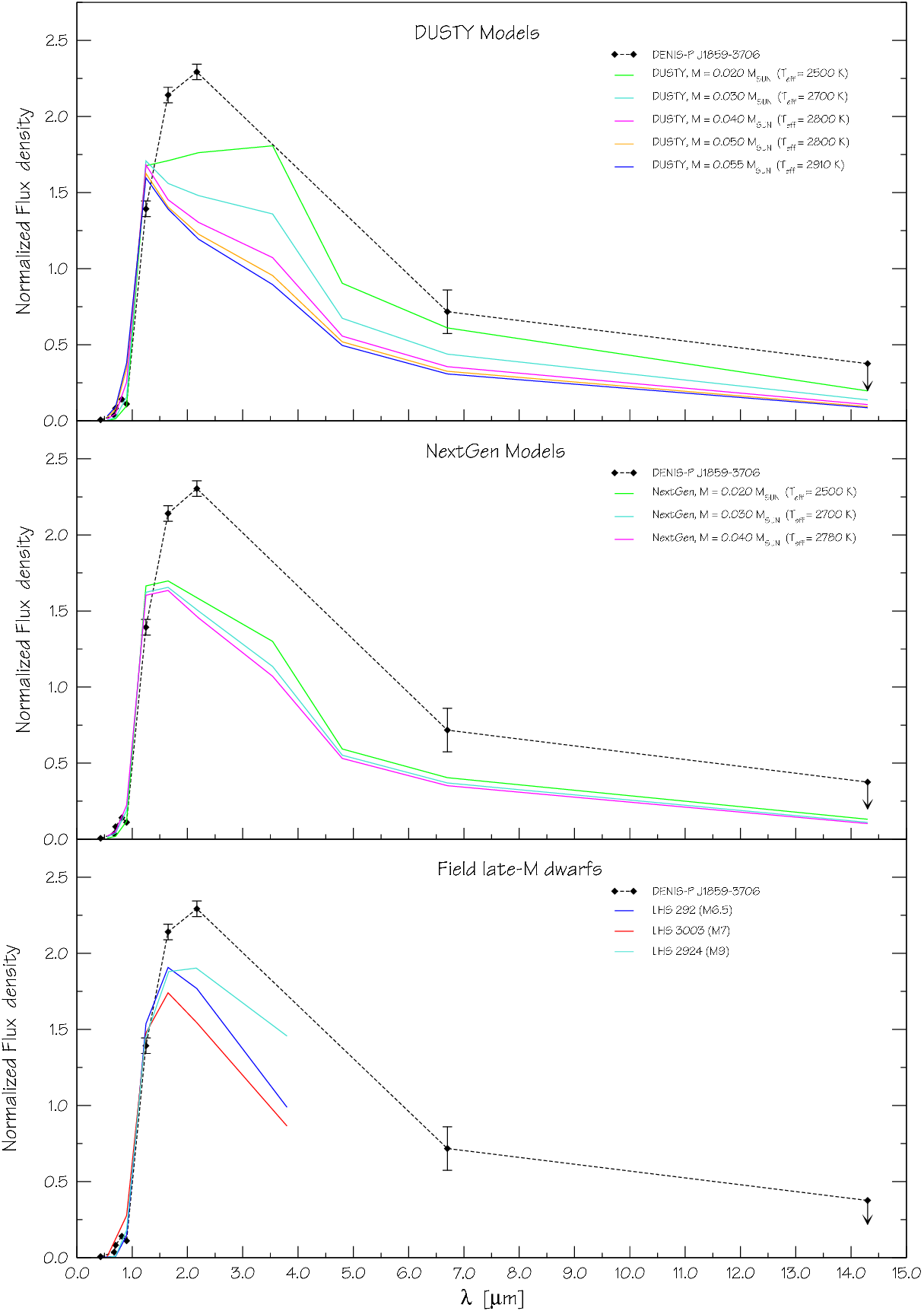}
      \caption{Spectral energy distribution of DENIS-P~J185950.9-370632 compared with DUSTY and NextGen models (5~Myr) and field late M-dwarfs distributions: \object{LHS~3003} (M7), \object{LHS~292} (M6.5), and \object{LHS~2924} (M9). The values for DENIS-P~J185950.9-370632 have been corrected for an extinction of A$_{V}$=0.5~mag. All fluxes have been normalized to their integrated luminosity between 0 and 1.65~$\mu$m. \citep[values for the field dwarfs from][]{1992ApJS...82..351L,2002ApJ...564..452L}.}
      \label{distri_spec}
   \end{figure*}

Figure \ref{distri_spec} compares the spectral energy distribution (SED) of DENIS-P~J185950.9-370632 to the DUSTY and NextGen models (for 5~Myr), as well as to observed field late-M dwarfs. The fluxes have been normalized to have an integrated luminosity between 0 and 1.65~$\mu$m (H band) equal to one. The choice of the H band as limit instead of, for example, the overall SED, was made for the following reasons: first because the SED of late-M dwarfs peaks around this value (see the DUSTY, NextGen and field M dwarfs SED in the figure), and second because in this wavelength range, the SED should be less affected by accretion-related continuum emission and infrared emission from the disk than at redder colours.

DENIS-P~J185950.9-370632 shows a strong excess over the models and the field dwarfs at wavelength greater than the H band. This wavelength range ($\lambda \ge$1.6~$\mu$m) is on the Raleigh-Jeans tail of the spectrum, and models are expected to be very reliable. We did not find any published observations of field late-M dwarfs in the 5.0~$\mu$m--8.5~$\mu$m range, but the available photometric measurements in the R to L' bands already show that the field M dwarfs have a much lower flux between K and L' than DENIS-P~J185950.9-370632, independently demonstrating the infrared excess. This excess refers to the unresolved system, which is probably dominated by the primary (see section \ref{binarity}). But since DENIS-P~J185950.9-370632 is a binary, we checked that its SED cannot be fitted by any meaningful combination of two synthetic SED, and the infrared excess is robust.

\subsection{An accreting close Binary}

As discussed in Section \ref{binarity}, DENIS-P~J185950.9-370632 is very likely to be a binary, with a separation of $\sim$0\farcs060, or 7.8~A.U at its 130~pc distance. This separation is well within the 0--20~A.U range observed for field brown dwarf systems \citep{2003AJ....126.1526B, 2003ApJ...586..512B, 2003ApJ...587..407C, 2003AJ....125.3302G}. The statistical correction factor of 1.26 from \citet{1992ApJ...396..178F}, leads to a semi-major axis of $a=$9.8~A.U. As discussed in section \ref{analysis}, the photometry, the spectral distribution, and the optical low resolution spectra all indicate an age of 5$\sim$10~Myr for a distance of 130~pc, and an effective temperature of $\sim$2\,600~K. 
The small magnitude difference (Table \ref{result_psf_fitting}) indicates that the two components of the system must have fairly similar masses. According to the DUSTY models, magnitude differences of $\sim$1.5~mag in the $F675W$ filter and $\sim$1.1~mag in the $F814W$ filter correspond to a mass ratio of $M_{B}/M_{A}\sim$75\%. Considering a total mass between 0.020~M$_{\sun}<$M$_{tot}<$0.030~M$_{\sun}$ and an orbit with a semi-major axis of 9.8~A.U, this leads to an orbital period of $\sim170$~yrs

\section{Conclusions}
The results presented in this paper demonstrate that DENIS-P~J185950.9-370632 is a young multiple system in the R--Cra star forming region. On the optical images, we find a clear elongation. After PSF subtraction on 3 epochs data we conclude that DENIS-P~J185950.9-370632 is very likely to be a common proper motion pair with a separation of $\sim$0\farcs060, close to  the resolution limit of the instruments we used. The spectroscopy constrains the effective temperature to 2600$\sim$2700~K for a low surface gravity (log $g$=3.5), consistent with a young age. This temperature corresponds to a total mass of $\sim$0.030$\pm$0.010~M$_{\sun}$ for an age ranging between 5 and 10~Myr. This mass is consistent with the photometry, and DENIS-P~J185950.9-370632 is therefore clearly substellar. Infrared excess observed  with the ISOCAM in the 5.0-8.5$\mu$m band  and the  presence of a strong H$\alpha$ emission as well as lithium absorption at  670.8~nm suggest a young age, the presence of circumstellar material, and  a sub-stellar mass object. Added to a consistent preliminary estimate of the  proper motion and consistent colours, these observations suggest that  DENIS-P~J185950.9-370632 belongs to the R--CrA star forming region. From the magnitude difference between the two components we estimate a mass ratio of $\sim$75\%. The estimated orbital period is about 170~yrs. The mid-infrared excess observed by ISO and the width of the H$\alpha$ emission suggest that it might be surrounded by a disk and slowly accreting. 

\begin{acknowledgements}
H. Bouy would like to thank M. Lombardi for computing the extinction map of the R--Cra region using the NICER technique \citep{2001A&A...377.1023L} and J. Alves and D. Apai for helpful discussions on this work.  The authors would like to thank the anonymous referee of this paper, whose precious comments helped to improve the quality of this article. This work is based on observations collected at the European Southern Observatory (Paranal, Chile), program 69.C-0588; and with the NASA/ESA Hubble Space Telescope, obtained at the Space Telescope Science Institute, which is operated by the Association of Universities for Research in Astronomy, Inc., under NASA contract NAS 5-26555. These observations are associated with programs GO8720 and GO9451. This work also uses archival data obtained with ISO, an ESA project with instruments funded by ESA Member States (especially the PI countries: France, Germany, the Netherlands and the United Kingdom) and with the participation of ISAS and NASA. This publication makes use of data from the DEep Near Infrared Survey. The DENIS project has been partly funded by the SCIENCE and the HCM plans of the European Commission under grants CT920791 and CT940627. It is supported by INSU, MEN and CNRS in France, by the State of Baden-Württemberg in Germany, by DGICYT in Spain, by CNR in Italy, by FFwFBWF in Austria, by FAPESP in Brazil, by OTKA grants F-4239 and F-013990 in Hungary, and by the ESO C\&EE grant A-04-046. This publication makes use of data products from the Two Micron All Sky Survey, which is a joint project of the University of Massachusetts and the Infrared Processing and Analysis Center/California Institute of Technology, funded by the National Aeronautics and Space Administration and the National Science Foundation. This work made use of data from the The Guide Star Catalogue~II, which is a joint project of the Space Telescope Science Institute and the Osservatorio Astronomico di Torino. 
\end{acknowledgements}

\newpage

   \begin{table*}

	      \caption[]{Astrometry} 
         \label{astrophotometry}
     \begin{tabular}{@{} lllllll @{}}
           \hline
	   \hline
Epoch           &  R.A          &      Dec.     &  Uncert.         &   Filter   & Mag.   &  Source \\
            \hline
01/01/1985	&  18 59 50.9	&  -37 06 31.0	&  $\pm$0\farcs1   &	R1       &  18.9  &  USNO-B1.0 \\
                &               &               &                  &    B2       &  21.7  & \\
                &               &               &                  &    R2       &  19.9  & \\
                &               &               &                  &    I        &  17.3  & \\
\hline
06/07/1989	&  18 59 50.9	&  -37 06 31.2	&  $\pm$0\farcs3   &	F       &  20.2   &  GSC2.21 \\
\hline
20/04/1996	&  18:59:50.7	&  -37:06:28.0	&  $\pm$6\farcs0   &    LW2     &  11.6  &  ISO \\
                &               &               &                  &    LW3     & $>$10.7 & \\
\hline
28/04/1999      &  18:59:50.9	&  -37:06:32.0	&  $\pm$1\farcs0   &	I       &  17.01 &  DENIS \\
                &               &               &                  &    J       &  13.99 & \\
                &               &               &                  &    K       &  12.60 & \\
\hline
25/06/1999	&  18:59:50.9	&  -37:06:31.3	&  $\pm$0\farcs1   &	J       &  13.98 &  2MASS \\
                &               &               &                  &    H       &  13.10 & \\
                &               &               &                  &    K       &  12.56 & \\
\hline
12/09/2000	&  18:59:50.9	&  -37:06:30.6	&  $\pm$1\farcs0   &	F675W   &  19.33 &  HST/WFPC2 \\
                &               &               &                  &    F814W   &  16.88 & \\
\hline
24/09/2002	&  18:59:50.8	&  -37:06:27.6	&  $\pm$1\farcs0   &	F625W   &        &  HST/ACS\\
                &               &               &                  &    F775W   &        & \\
                &               &               &                  &    F850LP  &        & \\
\hline
26/06/2003	&  18:59:50.8	&  -37:06:31.4	&  $\pm$1\farcs0   &  Long pass &  19.6  &  HST/STIS\\
            \hline
     \end{tabular}
   \end{table*}

   \begin{table}
	      \caption[]{Observation log.} 
         \label{observations}
     \begin{tabular}{@{} lllll @{}}
            \hline
	    \hline
	     \multicolumn{5}{c}{Imaging} \\
	    \hline
            Instrument     &  Filter    & Exp. & Date Obs. & Pixel Scale \\
	                   &            & Time [s]       &           & [\arcsec]  \\
            \hline
	    WFPC2-PC & F675W  & 600    & 19/09/2000 & 0\farcs045  \\
	    WFPC2-PC & F814W  & 300    & 19/09/2000 & 0\farcs045  \\
	    ACS-HRC  & F625W  & 1\,000 & 24/09/2002 & 0\farcs027  \\
	    ACS-HRC  & F775W  & 460    & 24/09/2002 & 0\farcs027  \\
	    ACS-HRC  & F850LP & 340    & 24/09/2002 & 0\farcs027  \\
	    ISO-CAM1 & LW2    & 4\,328 & 20/04/1996 & 6\farcs0    \\
	    ISO-CAM1 & LW3    & 4\,326 & 20/04/1996 & 6\farcs0    \\
            \hline

            \hline
	     \multicolumn{5}{c}{Spectroscopy} \\
	    \hline
            Instrument     &  Wavelength      & Exp.     & Date Obs. & Dispersion  \\
	                   &  Range [$\mu$m]  & Time [s] &           & [\AA/pixel]  \\
            \hline
	    VLT-FORS2     & 0.590-0.715 & 600 &  06/05/2002 & 0.60 \\
	    VLT-FORS2     & 0.690-0.910 & 200 &  06/05/2002 & 1.06 \\
	    HST-STIS      & 0.525-1.300 & 4140 & 16/06/2003  & 4.92  \\
	    Keck-HIRES    & 0.667-0.895 & 1200 & 30/05/2000 & 1.1   \\
            \hline
     \end{tabular}
   \end{table}

   \begin{table*}

	 \caption[]{PSF fitting results} 
         \label{result_psf_fitting}
         \begin{tabular}{llllll}
\hline
\hline
Date Obs.    &  Instr.   &  Filter   &  Sep.\footnotemark[1]   [\arcsec]           &  P.A\footnotemark[1] [\degr]          &  $\Delta$Mag\footnotemark[1]   \\
\hline
23/06/2003   &  STIS/CCD &  F28X50LP &  elongated\footnotemark[2]                   &  elongated\footnotemark[2]           & elongated\footnotemark[2]    \\
24/09/2002   &  ACS/HRC  &  F625W    &  0\farcs065$\pm$0\farcs001   &  279.2$\pm$0.1        & 1.74$\pm$0.06 \\
24/09/2002   &  ACS/HRC  &  F775W    &  0\farcs057$\pm$0\farcs0005  &  279.1$\pm$0.1        & 0.66$\pm$0.05 \\
24/09/2002   &  ACS/HRC  &  F850LP   &  elongated                  &  elongated           & elongated    \\
12/09/2000   &  WFPC2/PC &  F675W    &  0\farcs066$\pm$0\farcs003   &  283.8$\pm$1.2        & 1.30$\pm$0.11 \\
12/09/2000   &  WFPC2/PC &  F814W    &  0\farcs059$\pm$0\farcs003   &  271.8$\pm$1.2        & 1.10$\pm$0.11 \\
\hline
        \end{tabular}

\thanks{\footnotemark[1] The uncertainties reported here are 1-$\sigma$ uncertainties as explained in \citet{2003AJ....126.1526B}.\\}
\thanks{\footnotemark[2] As explained in the text, the PSF fitting program did not give good enough results, although the object is clearly elongated.}
\end{table*}

   \begin{table}
	 \caption[]{Spectral Features} 
         \label{spectral_features}
         \begin{tabular}{llll}
\hline
\hline
Feature      &  Instrument     &  $\lambda$ [nm]   &  EW [\AA]   \\
\hline
H$\alpha$    &  HST/STIS       &  656.3         &  -18  $\pm$ 3 \\
H$\alpha$    &  VLT/FORS2      &  656.3         &  -17  $\pm$ 2 \\
H$\alpha$    &  Keck/HIRES     &  656.3         &  -18  $\pm$ 3 \\
Na~\textsc{I}         &  VLT/FORS2      &  818.3         &  3.7 $\pm$ 0.3 \\
Li~\textsc{I}         &  VLT/FORS2      &  670.8         &  0.9 $\pm$ 0.4 \\
Li~\textsc{I}         &  Keck/HIRES     &  670.8         &  0.41$\pm$  0.08 \\
\hline
        \end{tabular}
\end{table}

\newpage


\bibliographystyle{aa}

\begin{thebibliography}{56}
\expandafter\ifx\csname natexlab\endcsname\relax\def\natexlab#1{#1}\fi

\bibitem[{{Adams} {et~al.}(1987){Adams}, {Lada}, \&
  {Shu}}]{1987ApJ...312..788A}
{Adams}, F.~C., {Lada}, C.~J., \& {Shu}, F.~H. 1987, \apj, 312, 788

\bibitem[{{Alard}(2000)}]{2000A&AS..144..363A}
{Alard}, C. 2000, \aaps, 144, 363

\bibitem[{{Allard} {et~al.}(2001){Allard}, {Hauschildt}, {Alexander},
  {Tamanai}, \& {Schweitzer}}]{2001ApJ...556..357A}
{Allard}, F., {Hauschildt}, P.~H., {Alexander}, D.~R., {Tamanai}, A., \&
  {Schweitzer}, A. 2001, \apj, 556, 357

\bibitem[{{Andre} \& {Montmerle}(1994)}]{1994ApJ...420..837A}
{Andre}, P. \& {Montmerle}, T. 1994, \apj, 420, 837

\bibitem[{{Baggett} {et~al.}(2002){Baggett}, {McMaster}, {Biretta}, {Burrows},
  {Casertano}, \& {Fruchteret}}]{WFPC2_DATA_HANDBOOK}
{Baggett}, S., {McMaster}, M., {Biretta}, J., {et~al.} 2002, HST WFPC2 Data
  Handbook, v. 4.0, ed. B. Mobasher, Baltimore, STScI

\bibitem[{{Barrado y Navascues} \& {Mart\'\i n}(2003)}]{Barrado...Martin}
{Barrado y Navascues}, D. \& {Mart\'\i n}, E.~L. 2003, \aj, in press

\bibitem[{{Basri} \& {Mart{\'{\i}}n}(1999)}]{1999AJ....118.2460B}
{Basri}, G. \& {Mart{\'{\i}}n}, E.~L. 1999, \aj, 118, 2460

\bibitem[{{Bate} {et~al.}(2002){Bate}, {Bonnell}, \&
  {Bromm}}]{2002MNRAS.332L..65B}
{Bate}, M.~R., {Bonnell}, I.~A., \& {Bromm}, V. 2002, \mnras, 332, L65

\bibitem[{{Blommaert} {et~al.}(2001){Blommaert}, {Siebenmorgen}, {Coulais},
  {Okumura}, {Ott}, {Sauvage}, \& {Starck}}]{ISOCAM..DATA..HANDBOOK}
{Blommaert}, J., {Siebenmorgen}, R., {Coulais}, A., {et~al.} 2001, ISO
  Handbook, Volume III

\bibitem[{{Bodenheimer}(1999)}]{Protostars..Planets..IV}
{Bodenheimer}, e.~a. 1999, Space Science Series: University of Arizona Press

\bibitem[{{Borsenberger}(1997)}]{DENIS_Reduction}
{Borsenberger}, J., e.~a. 1997, F. Garzon et al. (eds), Kluwer Dordrecht, 121

\bibitem[{{Bouy}(2004)}]{bouy2004}
{Bouy}, H. 2004, in prep.

\bibitem[{{Bouy} {et~al.}(2003){Bouy}, {Brandner}, {Mart{\'{\i}}n}, {Delfosse},
  {Allard}, \& {Basri}}]{2003AJ....126.1526B}
{Bouy}, H., {Brandner}, W., {Mart{\'{\i}}n}, E.~L., {et~al.} 2003, \aj, 126,
  1526

\bibitem[{{Burgasser} {et~al.}(2003){Burgasser}, {Kirkpatrick}, {Reid},
  {Brown}, {Miskey}, \& {Gizis}}]{2003ApJ...586..512B}
{Burgasser}, A.~J., {Kirkpatrick}, J.~D., {Reid}, I.~N., {et~al.} 2003, \apj,
  586, 512

\bibitem[{{Cambr{\' e}sy}(1999)}]{1999A&A...345..965C}
{Cambr{\' e}sy}, L. 1999, \aap, 345, 965

\bibitem[{{Casey} {et~al.}(1998){Casey}, {Mathieu}, {Vaz}, {Andersen}, \&
  {Suntzeff}}]{1998AJ....115.1617C}
{Casey}, B.~W., {Mathieu}, R.~D., {Vaz}, L.~P.~R., {Andersen}, J., \&
  {Suntzeff}, N.~B. 1998, \aj, 115, 1617

\bibitem[{{Chabrier} {et~al.}(2000){Chabrier}, {Baraffe}, {Allard}, \&
  {Hauschildt}}]{2000ApJ...542..464C}
{Chabrier}, G., {Baraffe}, I., {Allard}, F., \& {Hauschildt}, P. 2000, \apj,
  542, 464

\bibitem[{{Close} {et~al.}(2003){Close}, {Siegler}, {Freed}, \&
  {Biller}}]{2003ApJ...587..407C}
{Close}, L.~M., {Siegler}, N., {Freed}, M., \& {Biller}, B. 2003, \apj, 587,
  407

\bibitem[{{de Zeeuw} {et~al.}(1999){de Zeeuw}, {Hoogerwerf}, {de Bruijne},
  {Brown}, \& {Blaauw}}]{1999AJ....117..354D}
{de Zeeuw}, P.~T., {Hoogerwerf}, R., {de Bruijne}, J.~H.~J., {Brown}, A.~G.~A.,
  \& {Blaauw}, A. 1999, \aj, 117, 354

\bibitem[{{Delfosse} {et~al.}(2003){Delfosse}, {Martin}, {Forveille}, {Guieu},
  {Borsenberger}, {Epchtein}, {Fouque}, \& {Simon}}]{2003sf2a.confE.242D}
{Delfosse}, X., {Martin}, E.~L., {Forveille}, T., {et~al.} 2003, in SF2A-2003:
  Semaine de l'Astrophysique Francaise

\bibitem[{{Delgado-Donate} {et~al.}(2003){Delgado-Donate}, {Clarke}, \&
  {Bate}}]{2003MNRAS.342..926D}
{Delgado-Donate}, E.~J., {Clarke}, C.~J., \& {Bate}, M.~R. 2003, \mnras, 342,
  926

\bibitem[{{Epchtein} {et~al.}(1997){Epchtein}, {de Batz}, {Capoani},
  {Chevallier}, {Copet}, {Fouque}, {Lacombe}, {Le Bertre}, {Pau}, {Rouan},
  {Ruphy}, {Simon}, {Tiphene}, {Burton}, {Bertin}, {Deul}, {Habing},
  {Borsenberger}, {Dennefeld}, {Guglielmo}, {Loup}, {Mamon}, {Ng}, {Omont},
  {Provost}, {Renault}, {Tanguy}, {Kimeswenger}, {Kienel}, {Garzon}, {Persi},
  {Ferrari-Toniolo}, {Robin}, {Paturel}, {Vauglin}, {Forveille}, {Delfosse},
  {Hron}, {Schultheis}, {Appenzeller}, {Wagner}, {Balazs}, {Holl}, {Lepine},
  {Boscolo}, {Picazzio}, {Duc}, \& {Mennessier}}]{1997Msngr..87...27E}
{Epchtein}, N., {de Batz}, B., {Capoani}, L., {et~al.} 1997, The Messenger, 87,
  27

\bibitem[{{Fern{\' a}ndez} \& {Comer{\' o}n}(2001)}]{2001A&A...380..264F}
{Fern{\' a}ndez}, M. \& {Comer{\' o}n}, F. 2001, \aap, 380, 264

\bibitem[{{Fischer} \& {Marcy}(1992)}]{1992ApJ...396..178F}
{Fischer}, D.~A. \& {Marcy}, G.~W. 1992, \apj, 396, 178

\bibitem[{{Gizis} {et~al.}(2003){Gizis}, {Reid}, {Knapp}, {Liebert},
  {Kirkpatrick}, {Koerner}, \& {Burgasser}}]{2003AJ....125.3302G}
{Gizis}, J.~E., {Reid}, I.~N., {Knapp}, G.~R., {et~al.} 2003, \aj, 125, 3302

\bibitem[{{Hauschildt} {et~al.}(1999){Hauschildt}, {Allard}, \&
  {Baron}}]{1999ApJ...512..377H}
{Hauschildt}, P.~H., {Allard}, F., \& {Baron}, E. 1999, \apj, 512, 377

\bibitem[{{Jayawardhana} {et~al.}(2003){Jayawardhana}, {Ardila}, {Stelzer}, \&
  {Haisch}}]{2003AJ....126.1515J}
{Jayawardhana}, R., {Ardila}, D.~R., {Stelzer}, B., \& {Haisch}, K.~E. 2003,
  \aj, 126, 1515

\bibitem[{{Jayawardhana} {et~al.}(2002){Jayawardhana}, {Mohanty}, \&
  {Basri}}]{2002ApJ...578L.141J}
{Jayawardhana}, R., {Mohanty}, S., \& {Basri}, G. 2002, \apjl, 578, L141

\bibitem[{{Knacke} {et~al.}(1973){Knacke}, {Strom}, {Strom}, {Young}, \&
  {Kunkel}}]{1973ApJ...179..847K}
{Knacke}, R.~F., {Strom}, K.~M., {Strom}, S.~E., {Young}, E., \& {Kunkel}, W.
  1973, \apj, 179, 847

\bibitem[{{Leggett}(1992)}]{1992ApJS...82..351L}
{Leggett}, S.~K. 1992, \apjs, 82, 351

\bibitem[{{Leggett} {et~al.}(2002){Leggett}, {Golimowski}, {Fan}, {Geballe},
  {Knapp}, {Brinkmann}, {Csabai}, {Gunn}, {Hawley}, {Henry}, {Hindsley},
  {Ivezi{\' c}}, {Lupton}, {Pier}, {Schneider}, {Smith}, {Strauss}, {Uomoto},
  \& {York}}]{2002ApJ...564..452L}
{Leggett}, S.~K., {Golimowski}, D.~A., {Fan}, X., {et~al.} 2002, \apj, 564, 452

\bibitem[{{Lombardi} \& {Alves}(2001)}]{2001A&A...377.1023L}
{Lombardi}, M. \& {Alves}, J. 2001, \aap, 377, 1023

\bibitem[{{Mack} {et~al.}(2002){Mack}, {Boffi}, {Bohlin}, {Clampin}, {Cox},
  {DeMarchi}, {Ferguson}, {Gilliland}, {Hack}, {Hartig}, {Jedrzejewski},
  {Krist}, {Mutchler}, {O'Dea}, {Pavlovsky}, {Riess}, {Sparks}, {Stiavelli},
  {Suchkov}, {van Orsow}, {Welty}, {Tamanai}, \&
  {Schweitzer}}]{ACS_DATA_HANDBOOK}
{Mack}, J., {Boffi}, F., {Bohlin}, R., {et~al.} 2002, HST ACS Data Handbook,
  version 1.0, ed. B. Mobasher, Baltimore, STScI

\bibitem[{{Marraco} \& {Rydgren}(1981)}]{1981AJ.....86...62M}
{Marraco}, H.~G. \& {Rydgren}, A.~E. 1981, \aj, 86, 62

\bibitem[{{Mart{\'{\i}}n} {et~al.}(2003){Mart{\'{\i}}n}, {Barrado y Navascu{\'
  e}s}, {Baraffe}, {Bouy}, \& {Dahm}}]{2003ApJ...594..525M}
{Mart{\'{\i}}n}, E.~L., {Barrado y Navascu{\' e}s}, D., {Baraffe}, I., {Bouy},
  H., \& {Dahm}, S. 2003, \apj, 594, 525

\bibitem[{{Mart{\'{\i}}n} {et~al.}(1999){Mart{\'{\i}}n}, {Delfosse}, {Basri},
  {Goldman}, {Forveille}, \& {Zapatero Osorio}}]{1999AJ....118.2466M}
{Mart{\'{\i}}n}, E.~L., {Delfosse}, X., {Basri}, G., {et~al.} 1999, \aj, 118,
  2466

\bibitem[{{Mart{\'{\i}}n} {et~al.}(2004){Mart{\'{\i}}n}, {Delfosse}, \&
  {Guieu}}]{2004AJ....127..449M}
{Mart{\'{\i}}n}, E.~L., {Delfosse}, X., \& {Guieu}, S. 2004, \aj, 127, 449

\bibitem[{{Monet} {et~al.}(2003){Monet}, {Levine}, {Canzian}, {Ables}, {Bird},
  {Dahn}, {Guetter}, {Harris}, {Henden}, {Leggett}, {Levison}, {Luginbuhl},
  {Martini}, {Monet}, {Munn}, {Pier}, {Rhodes}, {Riepe}, {Sell}, {Stone},
  {Vrba}, {Walker}, {Westerhout}, {Brucato}, {Reid}, {Schoening}, {Hartley},
  {Read}, \& {Tritton}}]{2003AJ....125..984M}
{Monet}, D.~G., {Levine}, S.~E., {Canzian}, B., {et~al.} 2003, \aj, 125, 984

\bibitem[{{Muench} {et~al.}(2001){Muench}, {Alves}, {Lada}, \&
  {Lada}}]{2001ApJ...558L..51M}
{Muench}, A.~A., {Alves}, J., {Lada}, C.~J., \& {Lada}, E.~A. 2001, \apjl, 558,
  L51

\bibitem[{{Muzerolle} {et~al.}(2000){Muzerolle}, {Brice{\~ n}o}, {Calvet},
  {Hartmann}, {Hillenbrand}, \& {Gullbring}}]{2000ApJ...545L.141M}
{Muzerolle}, J., {Brice{\~ n}o}, C., {Calvet}, N., {et~al.} 2000, \apjl, 545,
  L141

\bibitem[{{Natta} \& {Testi}(2001)}]{2001A&A...376L..22N}
{Natta}, A. \& {Testi}, L. 2001, \aap, 376, L22

\bibitem[{{Neuh{\" a}user} {et~al.}(2000){Neuh{\" a}user}, {Walter}, {Covino},
  {Alcal{\' a}}, {Wolk}, {Frink}, {Guillout}, {Sterzik}, \& {Comer{\'
  o}n}}]{2000A&AS..146..323N}
{Neuh{\" a}user}, R., {Walter}, F.~M., {Covino}, E., {et~al.} 2000, \aaps, 146,
  323

\bibitem[{{Olofsson} {et~al.}(1999){Olofsson}, {Huldtgren}, {Kaas}, {Bontemps},
  {Nordh}, {Abergel}, {Andr{\' e}}, {Boulanger}, {Burgdorf}, {Casali},
  {Cesarsky}, {Davies}, {Falgarone}, {Montmerle}, {Perault}, {Persi}, {Prusti},
  {Puget}, \& {Sibille}}]{1999A&A...350..883}
{Olofsson}, G., {Huldtgren}, M., {Kaas}, A.~A., {et~al.} 1999, \aap, 350, 883

\bibitem[{{Papaloizou} \& {Terquem}(2001)}]{2001MNRAS.325..221P}
{Papaloizou}, J.~C.~B. \& {Terquem}, C. 2001, \mnras, 325, 221

\bibitem[{{Pascucci} {et~al.}(2003){Pascucci}, {Apai}, {Henning}, \&
  {Dullemond}}]{2003ApJ...590L.111P}
{Pascucci}, I., {Apai}, D., {Henning}, T., \& {Dullemond}, C.~P. 2003, \apjl,
  590, L111

\bibitem[{{Pinfield} {et~al.}(2003){Pinfield}, {Dobbie}, {Jameson}, {Steele},
  {Jones}, \& {Katsiyannis}}]{2003MNRAS.342.1241P}
{Pinfield}, D.~J., {Dobbie}, P.~D., {Jameson}, R.~F., {et~al.} 2003, \mnras,
  342, 1241

\bibitem[{{Preibisch} \& {Zinnecker}(1999)}]{1999AJ....117.2381P}
{Preibisch}, T. \& {Zinnecker}, H. 1999, \aj, 117, 2381

\bibitem[{{Reid} \& {Mahoney}(2000)}]{2000MNRAS.316..827R}
{Reid}, I.~N. \& {Mahoney}, S. 2000, \mnras, 316, 827

\bibitem[{{Reipurth} \& {Clarke}(2001)}]{2001AJ....122..432R}
{Reipurth}, B. \& {Clarke}, C. 2001, \aj, 122, 432

\bibitem[{{Shu} {et~al.}(1987){Shu}, {Adams}, \&
  {Lizano}}]{1987ARA&A..25...23S}
{Shu}, F.~H., {Adams}, F.~C., \& {Lizano}, S. 1987, \araa, 25, 23

\bibitem[{{Skrutskie}(1997)}]{2MASS}
{Skrutskie}, M.~F. 1997, The Impact of Large Scale Near IR Sky Surveys, ed. F.
  Garzon (Dordrecht: Kluwer), 25

\bibitem[{{Testi} {et~al.}(2002){Testi}, {Natta}, {Oliva}, {D'Antona},
  {Comeron}, {Baffa}, {Comoretto}, \& {Gennari}}]{2002ApJ...571L.155T}
{Testi}, L., {Natta}, A., {Oliva}, E., {et~al.} 2002, \apjl, 571, L155

\bibitem[{{Vogt} {et~al.}(1994){Vogt}, {Allen}, {Bigelow}, {Bresee}, {Brown},
  {Cantrall}, {Conrad}, {Couture}, {Delaney}, {Epps}, {Hilyard}, {Hilyard},
  {Horn}, {Jern}, {Kanto}, {Keane}, {Kibrick}, {Lewis}, {Osborne},
  {Pardeilhan}, {Pfister}, {Ricketts}, {Robinson}, {Stover}, {Tucker}, {Ward},
  \& {Wei}}]{1994SPIE.2198..362V}
{Vogt}, S.~S., {Allen}, S.~L., {Bigelow}, B.~C., {et~al.} 1994, in Proc. SPIE
  Instrumentation in Astronomy VIII, David L. Crawford; Eric R. Craine; Eds.,
  Volume 2198, p. 362, 362--+

\bibitem[{{Wilking} {et~al.}(1992){Wilking}, {Greene}, {Lada}, {Meyer}, \&
  {Young}}]{1992ApJ...397..520W}
{Wilking}, B.~A., {Greene}, T.~P., {Lada}, C.~J., {Meyer}, M.~R., \& {Young},
  E.~T. 1992, \apj, 397, 520

\bibitem[{{Wilking} {et~al.}(1997){Wilking}, {McCaughrean}, {Burton}, {Giblin},
  {Rayner}, \& {Zinnecker}}]{1997AJ....114.2029W}
{Wilking}, B.~A., {McCaughrean}, M.~J., {Burton}, M.~G., {et~al.} 1997, \aj,
  114, 2029

\bibitem[{{York}(2000)}]{2000AJ...129..1579}
{York}, D. G. e.~a. 2000, \aj, 129, 1579

\end{thebibliography}

\end{document}